\documentclass[aps,prb,twocolumn,amsmath,amssymb,superscriptaddress,nofootinbib]{revtex4-2}
\usepackage{titlesec}
\usepackage{graphicx}
\usepackage{subfigure}
\usepackage{mathrsfs}
\usepackage{amsthm}
\usepackage{amsmath}
\usepackage{float}
\usepackage{color}
\usepackage{braket}
\usepackage[pdftex, 
            pdfstartview=FitH,
            CJKbookmarks=true,
            bookmarksnumbered=true,
            bookmarksopen=true,
            colorlinks, 
            pdfborder=001,
            linkcolor=blue,
            anchorcolor=blue,
            citecolor=blue
            ]{hyperref}

\begin{document}

\title{Magnetic-flux tunable electronic transport through domain walls in a three-dimensional second-order topological insulator}

\author{Zhe Hou}
\email{zhe.hou@nnu.edu.cn}
\affiliation{School of Physics and Technology, Nanjing Normal University, Nanjing 210023, China}

\author{Ai-Min Guo}
\affiliation{Hunan Key Laboratory for Super-microstructure and Ultrafast Process, School of Physics, Central South University, Changsha 410083, China}

\begin{abstract}
The three-dimensional (3D) topological insulators (TIs), hosting topologically protected helical surface states, can be promoted into second-order TIs when a diagonal Zeeman term, typical of magnetic doping, is introduced. The latter hosts exotic chiral one-dimensional (1D) topological hinge states (THSs). In this paper, we investigate the electronic transport of THSs through a magnetic domain wall (DW) in a 3D TI nanowire. Due to the sign reversal of the out-of-plane magnetization across the DW, four 1D topological boundary states, residing on the edge of the DW, arise and form an enclosed loop mediating the counterpropagating THSs. By applying a uniform magnetic field parallel to the nanowire, we obtain a perfect sinusoidal Aharonov-Bohm oscillation in the two-terminal conductance $G$, formulated by $G=\frac{e^2}{2h} \left[ 1- \cos(\pi \Phi/\Phi_0) \right]$, with $\Phi$ the magnetic flux through the DW and $\Phi_0 = h/2e$ the flux quantum. Applying a phenomenological scattering matrix approach, we explain this novel Aharonov-Bohm oscillation perfectly, and attribute the constructive (destructive) interference of transmission at $\Phi = \Phi_0$ (0) to the $\pi$-spin rotation of the THSs traveling through the DW. Extending our study to a double-DW junction, where the central region has antiparallel magnetization to the leads, we observe Fabry-P{\'e}rot oscillations, in which the conductance minima are tuned by the magnetic flux. Our findings open a new avenue for finely controlling the quantum transport of THSs in magnetic systems using magnetic flux, and provide a faithful way for detecting THSs in experiments.

\end{abstract}

\maketitle

\section{Introduction} 
Magnetic domain walls (DWs) or domain junctions exist ubiquitiously in magnetic nanostructures and play an essential role in various applications in condensed matter physics. One famous example is the giant magnetoresistance effect \cite{Baibich1988Giant, Binasch1989Enhanced, Berkowitz1992Giant, Xiao1992Giant, Fert1995Giant, Ennen2016Giant} in ferromagnetic materials, where the electrical resistance is relatively high (low) when the magnetization of the adjacent ferromagnetic layers is in an antiparallel (parallel) alignment. This effect arises from spin-dependent electronic transmission, and underpins the design of magnetoresistance sensors, a crucial component of hard disk drives. With the advent of topological insulators (TIs) \cite{Kane2005Quantum, Kane2005Topological, Fu2007Topological, Bernevig2006Quantum, Konig2007Quantum}, the combination of magnetism and topology has led to abundant phenomena in nanoelectronics \cite{Liu2008Quantum, Yu2010Quantized, Chang2013Experimental, Bestwick2015Precise, Shamim2021Quantized, Hor2010Development, Xu2012Hedgehog, Rienks2019Large, Nomura2010Electric, Checkelsky2012Dirac, Wakatsuki2015Domain, Yasuda2017Quantized, Zhou2018Configuration, Sedlmayr2020Chiral, Yang2020Linear}. For example, in Cr-doped thin films of three-dimensional (3D) TIs, the quantum anomalous Hall insulator (QAHI) phase arises with a vanishing longitudinal resistance and a quantized Hall resistance of value $h/e^2$ \cite{Chang2013Experimental, Bestwick2015Precise}. By constructing a DW structure in the QAHI, two new chiral edge states appear on the boundary of two domains with opposite Chern numbers \cite{Yasuda2017Quantized, Zhou2018Configuration, Sedlmayr2020Chiral, Yang2020Linear}. Such edge states are shown to be tunable by flipping the magnetization direction using a tip of the magnetic force microscope \cite{Yasuda2017Quantized}. In addition, the quantum transport mediated by the chiral edge states on the DW can also be flexibly engineered by tuning the width of the DW or the magnetization configuration \cite{Zhou2018Configuration, Yang2020Linear}, providing a convenient way to design new low-power-consumption electronics with magnetic TIs.

The concept of topology has been generalized to higher-order topological insulators (HOTIs) \cite{Benalcazar2017Quantized, Benalcazar2017Electric, Li2021Third, Schindler2018Higher, Schindler2018Bismuth, Noguchi2021Evidence, Langbehn2017Reflection, Ren2020Engineering, Hou2023Realization,  Geier2018Second, Song2017Dimensional, Weber2023Second, Wang2021Structural, Levitan2020Second, Queiroz2019Splitting, Luo2021Aharonov, Li2021Higher, Chaou2023Hinge}, where a new bulk-edge correspondence has been established. For an $n$-th order HOTI with dimension $\mathcal{N}$, it features $(\mathcal{N}-n)$-dimensional topological states on its surface or boundary. For instance, a two-dimensional (2D) second-order topological insulator (SOTI) exhibits localized zero-dimensional topological corner states \cite{Benalcazar2017Quantized, Ren2020Engineering, Song2017Dimensional, Weber2023Second}. For a 3D SOTI, it hosts linearly dispersing one-dimensional (1D) topological states propagating along the system's hinges, commonly referred to as the topological hinge states (THSs) \cite{Schindler2018Higher, Schindler2018Bismuth, Noguchi2021Evidence, Langbehn2017Reflection, Hou2023Realization, Geier2018Second}. 

\begin{figure}
\includegraphics[width=8.5cm, clip=]{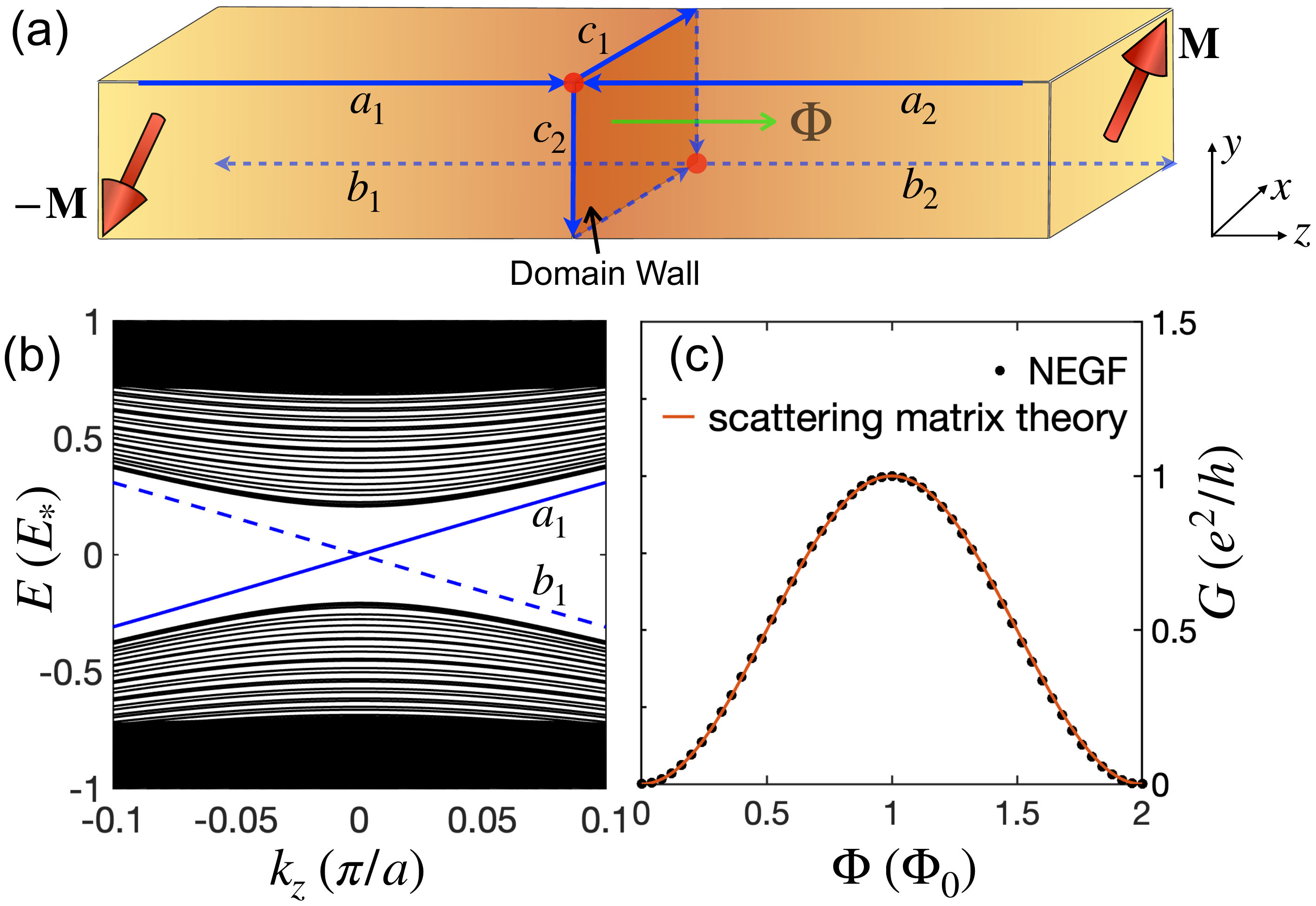} 
\caption{(a) Schematic diagram of the magnetic SOTI transport device, featuring a DW at the interface of two magnetic domains with antiparallel magnetization (indicated by the 3D red arrows). The blue arrows represent the propagation paths of the THSs, with $a_{1(2)}$, $b_{1(2)}$ denoting the THSs, and $c_{1(2)}$ denoting two of the four topological boundary states localized at the DW edge. A uniform magnetic field is applied along the $z$-direction, generating a magnetic flux $\Phi$ threading through the DW (see the green arrow). (b) Band structure of an infinitely-long SOTI nanowire with magnetization along the $(-1,-1,0)$ direction. Here the flux $\Phi = 0$. The two linearly dispersing bands correspond to $a_1$ and $b_1$ THSs in (a). (c) Two-terminal conductance $G$ through the device in (a), plotted as a function of magnetic flux ${\Phi}$ at Fermi energy $E_F = 0.02 E_*$. Black dots represent the numerical results obtained via the NEGF method, while the red curve shows the fitting using the scattering matrix approach. Throughout the paper, the magnetization strength is set to $M = 0.3 E_*$, and the DW region has a lateral size of $N_x=N_y=25$. }
\label{fig.SingleDWDevice}
\end{figure}

A practical route to realize 3D SOTI is introducing a diagonal effective Zeeman term \cite{Langbehn2017Reflection, Ren2020Engineering, Hou2023Realization} into the 3DTI through magnetic doping. This generates THSs at the interfaces of adjacent surfaces with opposite out-of-plane magnetizations [see Fig. \ref{fig.SingleDWDevice}(a) the $a_{1(2)}$ modes]. If the magnetization orientation is partially flipped, one naturally obtains a magnetic DW junction [see Fig. \ref{fig.SingleDWDevice}(a)]. This should happen inevitably in realistic magnetic materials, where magnetic domains exist ubiquitously during large-scale fabrication processes. However, to date, the investigation on the influence of magnetic DWs on the electronic transport in SOTIs—especially on the propagation of THSs, is still lacking. 

In this paper, we address this problem by constructing a magnetic DW structure in a 3D SOTI [see Fig. \ref{fig.SingleDWDevice}(a)], and systematically investigate its electronic transport properties. By introducing antiparallel magnetic doping on the two sides of a 3DTI nanowire, a DW forms at the interface. The sign reversal of the out-of-plane magnetization across the DW gives rise to four 1D topological boundary states appearing on the DW edges [see $c_1, c_2$ modes in Fig. \ref{fig.SingleDWDevice}(a) for two representative modes]. These boundary states mediate the propagation of the THSs, forming an enclosed loop for the electron's trajectory. By applying a uniform magnetic field perpendicular to the DW, we obtain a perfect sinusoidal Aharonov-Bohm (AB) oscillation in the two-terminal conductance [see Fig. \ref{fig.SingleDWDevice}(c)], which can be formulated by $G=\frac{e^2}{h} T$, with $T = 0.5 \left[ 1- \cos(\pi \Phi/\Phi_0) \right]$, where $\Phi$ is the magnetic flux threading the DW, and $\Phi_0 = h/2e$ is the flux quantum. This result indicates that a perfect transmission with quantized conductance $G=e^2/h$ occurs when $\Phi=\Phi_0$, while transmission is completely suppressed at $\Phi=0$. Using a phenomelogical scattering matrix approach, we successfully explain this novel AB oscillation, and identify the origin of the constructive (destructive) interference of transmission at $\Phi = \Phi_0$ (0) as the $\pi$-spin rotation of the THSs as they traverse the DW. 

We next consider a double-DW junction, where a central cavity region with magnetization antiparallel to that of the outer two leads is formed. Due to interference of two THSs within the cavity, a Fabry-P{\'e}rot (FP) oscillation for conductance is obtained, as the Fermi energy or the cavity length is varied. Unlike the single DW case, in the double-DW junction it is the minimum value of conductance $G_{DB}^{min}$ that is tuned by the magnetic flux, following the relation: $G_{DB}^{min} = \frac{e^2}{h} T^2/(2-T)^2$. This indicates that the transmission through the double-DW junction can still be switched on/off by tuning the flux to $\Phi_0$ or zero. Again the results can be perfectly explained by the scattering matrix approach. In view of the potential application of THSs in the realization of functional electronics and spintronics, our work opens a new avenue for controllable quantum transport in magnetic SOTIs via magnetic flux, and offers a robust experimental approach for identifying and characterizing THSs.

This paper is organized as follows: in Sec. \ref{sec: ModelMethods} we introduce the Hamiltonian for the 3D SOTI, and the non-equilibrium Green's function (NEGF) method in calculating the conductance. In Sec. \ref{sec: SingleDW}, we show the AB oscillation of conductance in a single-DW junction, and develop phenomenological scattering matrices to describe the propagation of THSs. In Sec. \ref{sec: DoubleDWs} we extend our analysis to a double-DW junction and demonstrate the emergence of FP oscillation. Sec. \ref{sec: Conclusion} is a concluding remark. Additional supplementary materials are presented in the Appendix.

\section{Model and methods \label{sec: ModelMethods}}
\subsection{SOTI Hamiltonian}
We start from a 3DTI model on a cubic lattice harboring both spin and orbital degrees of freedom  \cite{Hou2023Realization, Hosur2010Chiral, Hosur2011Majorana, Shen2013Topological}, whose Hamiltonian reads:  
\begin{align}
\hat{H}_0 = \sum_{\bf k}  \Psi^\dagger_{\bf k} H_0({\bf k}) \Psi_{\bf k},
\end{align}
where 
\begin{align}
\Psi^\dagger_{\bf k} = [\Psi^\dagger_{1 \uparrow, {\bf k}}, \Psi^\dagger_{1 \downarrow, {\bf k}}, \Psi^\dagger_{2 \uparrow, {\bf k}}, \Psi^\dagger_{2 \downarrow, {\bf k}}]
\end{align}
is the creation operator at wave vector $\bf k$, with subscripts $\uparrow$ ($\downarrow$) and 1 (2) referring to electron's spin and orbitals, respectively. Explicitly, $H_0({\bf k})$ can be written into: 
\begin{align}
H_0({\bf k}) =& \frac{ \hbar v}{a} \sigma_x \otimes (s_x \sin{k_x a} + s_y \sin{k_ya} + s_z \sin{k_z a}) + \nonumber \\
& \sigma_z \left[m_0-m_1(\cos{k_x a}+\cos{k_y a} + \cos{k_z a} ) \right],
\end{align}
where $\hbar$ is the reduced Planck's constant, $v$ denotes the Fermi velocity, $a$ is the lattice constant, and $m_{0, 1}$ are the model parameters which decides the topology of $H_0({\bf k})$ \cite{Hosur2011Majorana, Shen2013Topological}. The Pauli matrices $\bm \sigma$ and $\bf s$ act on the orbital and spin spaces, respectively. 

Next we introduce an effective Zeeman effect, as can be generated by magnetic doping of Mn or Cr atoms onto 3DTI \cite{Chang2013Experimental, Bestwick2015Precise, Hor2010Development, Xu2012Hedgehog, Rienks2019Large}, to induce a SOTI phase. The Hamiltonian now becomes $\hat{H}_1 = \sum_{\bf k} \Psi^\dagger_{\bf k} H_1({\bf k})  \Psi_{\bf k}$ with
\begin{align}
H_1({\bf k}) = H_0({\bf k}) +   {\bf M} \cdot {\bf s}.
\end{align}
Hereafter we fix the magnetization orientation along $\pm {\bf n}_{110}$ with ${\bf n}_{110} = \frac{1}{\sqrt{2}}(1, 1, 0)$, i.e., the diagonal direction in the $x-y$ plane, so that two 1D THSs appear if one considers an infinitely-long TI nanowire in a cubic geometry. The rise of THSs is attributed to the sign reversal of the out-of-plane magnetization on adjacent facets, akin to the Jackiw-Rebbi zero modes \cite{Jackiw1976Solitons}. 

Before proceeding, we shall discuss the symmetry of the Hamiltonian $\hat{H}_1$. First, the time-reversal symmetry with operator $\hat{T} = -i s_y \hat{K}$, where $\hat{K}$ is the complex conjugation, and the chiral symmetry with operator $\hat{S} = \sigma_y$, are both explicitly broken due to the presence of the Zeeman term ${\bf M} \cdot {\bf s}$. The inversion symmetry with $\hat{\Pi} = \sigma_z \hat{P}_{\bf r}$, where $\hat{P}_{\bf r}$ denotes the parity transformation ${\bf r} \rightarrow - {\bf r}$, is still preserved obeying the relation:
\begin{align}
\sigma_z H_1({\bf k}) \sigma_z = H_1(-{\bf k}).
\end{align}
One can also find that the particle-hole symmetry $\hat{P} = \sigma_y \otimes s_y \hat{K}$ is preserved by checking the following relation:
\begin{align}
\sigma_y \otimes s_y H^*_1({\bf k}) \sigma_y \otimes s_y = -H_1(-{\bf k})
\end{align} 
The combination of inversion-symmetry and particle-hole symmetry yields:
\begin{align}
(\sigma_z \hat{P}) H_1({\bf k}) (\sigma_z \hat{P})^{-1} = - H_1({\bf k}),
\end{align}
This indicates a particle-hole symmetry for the band structure of $H_1 ({\bf k})$. In Fig. \ref{fig.SingleDWDevice}(b) we plot the band structure by considering a SOTI nanowire with magnetization ${\bf M}$ along $-{\bf n}_{110}$ direction, where indeed we find this particle-hole symmetry. Besides, two linearly dispersing modes shown in blue solid and dashed lines appear, corresponding to two THSs as shown with $a_1$ and $b_1$ in Fig. \ref{fig.SingleDWDevice}(a). These two THSs are related by the inversion transformation. 

Finally, since the Zeeman term is applied along the $\pm {\bf n}_{110}$ direction, one can also define a mirror symmetry with $\hat{M}_{110} = i \sigma_z \otimes s_{xy} \hat{P}_{110}$. Here $s_{xy} = {\bf s} \cdot {\bf n}_{110} = \frac{1}{\sqrt{2}}(s_x + s_y)$, and $\hat{P}_{110}$ denotes the mirror transformation about the (110) plane: $x\rightarrow -y$, $y \rightarrow -x$, and $z \rightarrow z$. This symmetry relation reads:
\begin{align}
\sigma_z \otimes s_{xy} H_1 (k_x, k_y, k_z) \sigma_z \otimes s_{xy} = H_1 (-k_y, -k_x, k_z).
\end{align}
Since $\hat{M}_{110}$ commutes with $\hat{H}_1$, we can always find a simultaneous eigenstate for $\hat{M}$ and $\hat{H}_1$. This indicates that the THSs, which are the eigenstates of $\hat{H}_1$ in a cubic nanowire geometry, should only appear on the hinges where the mirror plane and the nanowire cross \cite{Langbehn2017Reflection}. This is consistent with the numerical calculations on THSs shown in Figs. \ref{fig.SingleDWDevice}(a) and \ref{fig.SingleDWDevice}(b).

\subsection{Tight-binding Hamiltonian and non-equilibrium Green's function method for calculating the conductance}

Next we construct a device with a DW structure [see Fig. \ref{fig.SingleDWDevice}(a)], to calculate the electronic transport properties of the SOTI. The DW is generated from the antiparallel magnetization on the two leads, and can be expressed by:
\begin{eqnarray}
{\bf M}({\bf r})=\left\{
\begin{array}{ll}
 -M {\bf n}_{110}   \hspace{1mm} & { \rm for}   \   {z      <   0  }\\
  M {\bf n}_{110}   \hspace{1mm} & {\rm for}  \  {z     \geq   0 },
\end{array} \right.
\end{eqnarray} where $M>0$ characterizes the strength of the magnetization. Here a sharp DW configuration is considered (see Appendix A for the discussion on the DW configurations).
In this case, another sign reversal of the out-of-plane magnetization across the DW happens, generating four 1D topological boundary states on the DW edges [see $c_1$ and $c_2$ modes shown in solid arrows in Fig. \ref{fig.SingleDWDevice}(a) as examples]. For the right-hand side of the device, the THSs, denoted by $a_2$ and $b_2$, appear at the same hinges as the left one, but host opposite chirality. Those counter-propagating modes are connected by the topological boundary states, forming an enclosed loop at the DW (see Appendix B for the distribution of the local density of states at the DW). This provides a platform to induce the AB oscillation if a magnetic flux $\Phi$ is threaded through the DW. We then apply a uniform magnetic field along the $z$-direction, i.e., the transport direction. To model this system, we discretize the Hamiltonian into a real-space tight-binding form:
\begin{align}
\label{eq. HDW}
\hat{H}_{DW}=&  \sum_l  d^\dagger_l  \mathcal{E}_l d_l \\ \nonumber
 &+ \sum_{ l,\ \alpha=x, y, z} [ d^\dagger_l T_\alpha e^{i\phi_{l, l+\alpha}}    
 d_{l+\alpha} +h.c. ],
\end{align}
where
\begin{align}
d^\dagger_l = \frac{1}{\sqrt{N_s}} \sum_{\bf k} e^{-i {\bf k} \cdot {\bf r}_l} \Psi^\dagger_{\bf k}
\end{align}
is the creation operator at site $l$, with ${\bf r}_l$ the real-space coordinate of site $l$, and $N_s$ the number of total sites. The on-site and hopping terms are: $\mathcal{E}_l=m_0\sigma_z +  {\bf M} ({\bf r}_l) \cdot {\bf s}$ and $T_\alpha=\frac{\hbar v }{2ia}\sigma_x \otimes  s_\alpha-\frac{m_1 }{2}\sigma_z \otimes s_0$. The orbital effect of the magnetic field is incorporated by adding a phase: $\phi_{l, l+\alpha}= \frac{e}{\hbar} \int_{{\bf r}_l}^{{\bf r}_l+{\bf e}_\alpha} {\bf A}\cdot d {\bf r}$ into the hopping matrix $T_\alpha$, with ${\bf e}_\alpha$ the unit vector pointing along the $\alpha$-direction, and ${\bf A} = (-B y, 0, 0)$ the vector potential. The Zeeman effect of the magnetic field has been omitted since experimentally the area of the DW can be made large so that the magnetic field needed in inducing one flux quantum is rather weak. 

We use the NEGF method to calculate the two-terminal conductance.  First we obtain the surface Green's function ${\bf g}^r_{L(R)}(E)$ of the left(right) terminal using the recursive Green's function method \cite{Surface_GF}, with $E$ the energy. The self-energy coupled to the central region is thus calculated to be: ${\bm \Sigma}^r_{L(R)}(E)={\bf H}_{cL(R)}{\bf g}^r_{L(R)}(E) {\bf H}^\dagger_{cL(R)}$ with ${\bf H}_{cL(R)}$ the coupling matrix from the central region to the left(right) terminal. Then the retarded Green's function of the central region is: ${\bf G}^r_c(E)=\left[ (E+i 0^+){\bf I}-{\bf H}_c-{\bm \Sigma}^r_L-{\bm \Sigma}^r_R \right]^{-1}$, with ${\bf H}_c$ the Hamiltonian of the central region. The transmission coefficient $T(E)$ is thus calculated by \cite{Meir1992LandauerFormula, Jauho1994TransportResonant, Datta1995Mesoscopic}:
\begin{eqnarray}
T(E)={\rm Tr}\left[ {\bm \Gamma}_R {\bf G}^r_c {\bm \Gamma}_L {\bf G}^a_c \right],
\end{eqnarray}
where ${\bm \Gamma}_{L(R)}(E)\equiv i \left[ {\bm \Sigma}^r_{L(R)}-({\bm \Sigma}^r_{L(R)})^\dagger \right]$ is the linewidth function for terminal L(R) and ${\bf G}^a_c(E)=\left[ {\bf G}^r_c(E) \right]^\dagger$ is the advanced Green's function of the central region. The differential conductance at Fermi energy $E_F$ at zero temperature is: $G(E_F)= T(E_F) \frac{e^2} h$.

To fit with realistic material parameters in \emph{ab initio} calculations \cite{Zhang2009Topological, Liu2010Model, Qi2011Topological} or experimental observations \cite{Xia2009Observation, Chen2009Experimental} on $\rm Bi_2Se_3$ or $\rm Bi_2Te_3$, we set the Fermi velocity $v=5 \times 10^5 \ {\rm m/s}$ and the lattice constant $a=2.2 \ {\rm nm}$. In the following numerical calculations, we define $E_*=\hbar v/a \approx 150 \ {\rm meV}$ as the energy unit. By choosing $m_0=2$ and $m_1=1$ (in units of $E_*$), we identify a strong TI phase for $\hat{H}_0 $ which features one massless Dirac-fermion excitation on each facet. Experimentally the surface gap in Mn-doped Bi$_2$Te$_3$ has been reported to be as large as 90 meV \cite{Rienks2019Large}. Furthermore, in the intrinsic magnetic TI MnBi$_2$Te$_4$ \cite{Otrokov2019Prediction, Zhang2019Topological, Sun2020Analytical, Choi2023Conductance, Sun2023Magnetic, Chen2024Layer}---an interlayer antiferromagnet that hosts exotic topological axion states \cite{Zhang2019Topological} and layer-selective quantum transport \cite{Choi2023Conductance, Sun2023Magnetic, Chen2024Layer}, the surface gap of about 70 meV has been experimentally observed at the (0001) surface \cite{Otrokov2019Prediction}. So here we set the magnetization strength to $M = 0.3 E_*$ ($\sim$45 meV). The bulk gap $\Delta_{bulk}$ opened in our model is thus estimated to be $1.4 E_*$  ($\sim$210 meV), and the surface gap $\Delta_{sur} \approx 0.4 E_*$ ($\sim$62 meV), as can be read from the band structure in Fig. \ref{fig.SingleDWDevice}(b). During numerical calculations, we discretize the cross-section of the transport device into $N_x \times N_y$ sites in the $x-y$ plane with open boundary condition. The magnetic flux threading through the DW is defined as: $\Phi = (N_x-1)(N_y-1) B a^2$. A square shape of the DW is adopted, so the transport system respects the mirror symmetry $\hat{M}_{110}$ when ${\Phi} = 0$. Throughout the paper we set $N_x=N_y=25$ during numerical calculations.

\section{Aharonov-Bohm oscillation of conductance through a single domain wall \label{sec: SingleDW}}
\subsection{Numerical results for conductance}

\begin{figure}
\includegraphics[width=8.5cm, clip=]{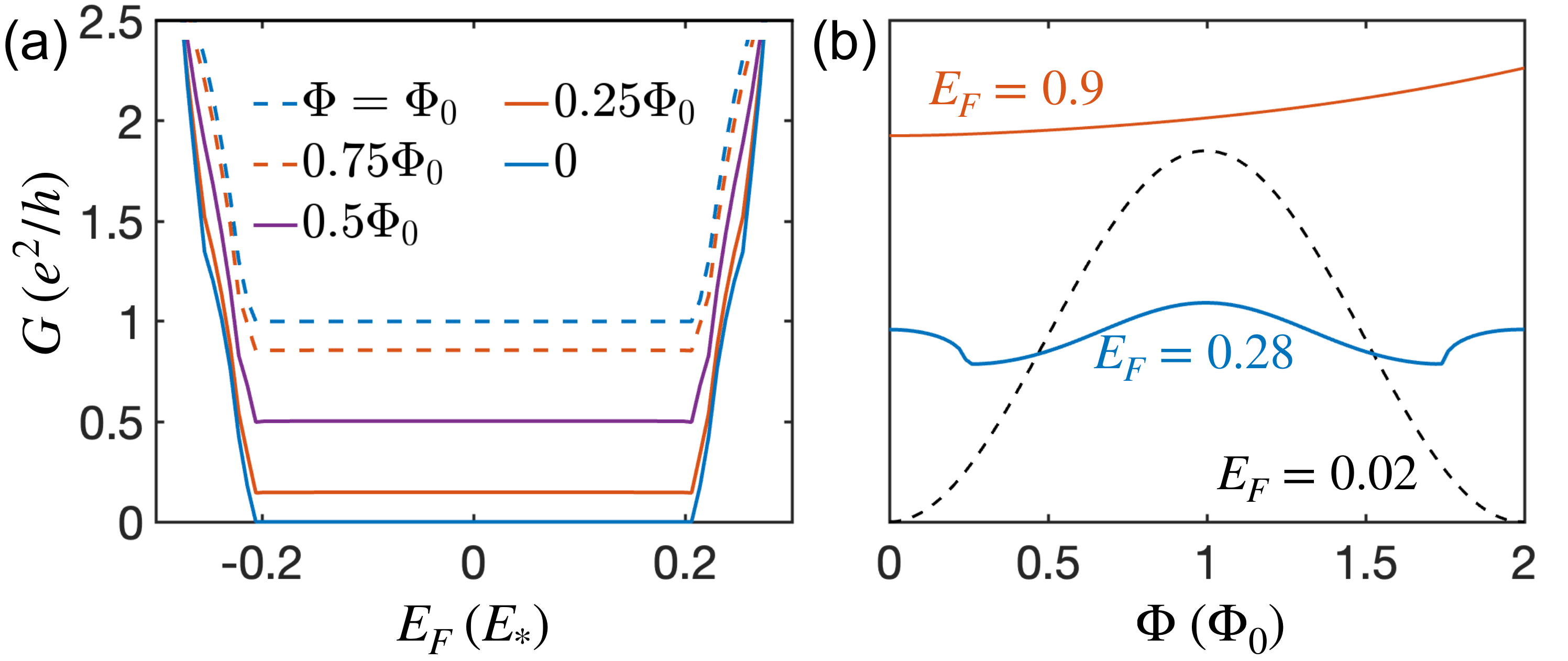} 
\caption{Conductance $G$ through a single DW, as a function of the Fermi energy $E_F$ with different magnetic flux ${\Phi}$ in (a), and as a function of the flux ${\Phi}$ at different Fermi energies (in units of $E_*$) in (b). In (b) the curves have been vertically offset for comparison. }
\label{fig.BulkConductance}
\end{figure}

Figure \ref{fig.SingleDWDevice}(c) shows the numerical calculations on conductance $G$ through a single DW using the NEGF method. Here the Fermi energy is chosen at $E_F = 0.02E_*$, which lies inside the surface gap $\Delta_{surf}$ and crosses with two THSs $a_1$, $b_1$ in Fig. \ref{fig.SingleDWDevice}(b). As the magnetic flux ${\Phi}$ varies from zero to $2 \Phi_0$, a perfect AB conductance oscillation is obtained, due to the formation of an enclosed loop at the DW. To investigate the Fermi energy $E_F$ dependence of this AB oscillation, in Fig. \ref{fig.BulkConductance}(a) we fix the flux ${\Phi}$ and then scan the Fermi energy $E_F$ for each curve. The conductance remains invariant as long as $E_F$ lies inside the surface gap, while increases sharply once the $E_F$ exceeds the surface gap and crosses with the surface states. In Fig. \ref{fig.BulkConductance}(b) we select three Fermi energies: $E_F = 0.02 E_*$, $0.28 E_*$, and $0.9 E_*$, which correspond to the 1D THSs, the 2D surface states, and the 3D bulk states, respectively, and plot their $G( {\Phi} )$ curves. As can be seen, the AB oscillation disappears once $E_F$ exceeds the surface gap, i.e., $|E_F| > \Delta_{surf}/2 \approx 0.2 E_*$. This disappearance can be understood from two aspects: I) the number of incoming (outgoing) modes from surface or bulk states is more than 1, resulting in a more complicated interference pattern; II) unlike the THSs which are unaffected by the magnetic field due to their 1D nature, the 2D surface states or 3D bulk states on the leads have velocity components perpendicular to the magnetic flux, and get strongly  influenced by the magnetic field due to the magnetic flux or the Lorentz force. 

The two THSs on the opposite hinges become coupled when the system sizes $N_x$ and $N_y$ are reduced. This coupling opens a small energy gap $\Delta_{coup}$, similar to the gap opening by the coupling between surface states in 3DTIs \cite{Zhang2010Crossover, Lin2023Tuning}. In this case, both the conductance plateaus shown in Fig. \ref{fig.BulkConductance}(a) and the AB oscillation in Fig. \ref{fig.BulkConductance}(b) are destroyed once the Fermi energy lies within this gap. To avoid this, the system size should be much larger than the localization length $\xi$ of the THSs. To estimate $\xi$, we use the scaling relation of the coupling-induced gap: $\Delta_{coup} \propto e^{-r/\xi}$ where $r$ is the distance between the two THSs. Since the wavefunctions of the THSs decay exponentially on the surface rather than the bulk (see Appendix B the local density of states), we can approximate $r \approx (N_x + N_y)a$. By analyzing the size dependence of $\Delta_{coup}$, we extract $\xi \approx 4.7a$ from an exponential fit. This value is very small compared with the system size we choose here, so the coupling between the two THSs can be well avoided in the above numerical calculations.

The numerical results here suggest that the observed sinusoidal AB oscillations are a distinctive feature of THSs. Given that these 1D THSs are embedded within the bulk of 3D SOTIs—making them challenging to detect using conventional techniques such as scanning tunneling spectroscopy—the AB oscillations presented here offer a powerful and unambiguous probe for identifying and validating the presence of THSs.

\begin{figure}
\includegraphics[width=8.5cm, clip=]{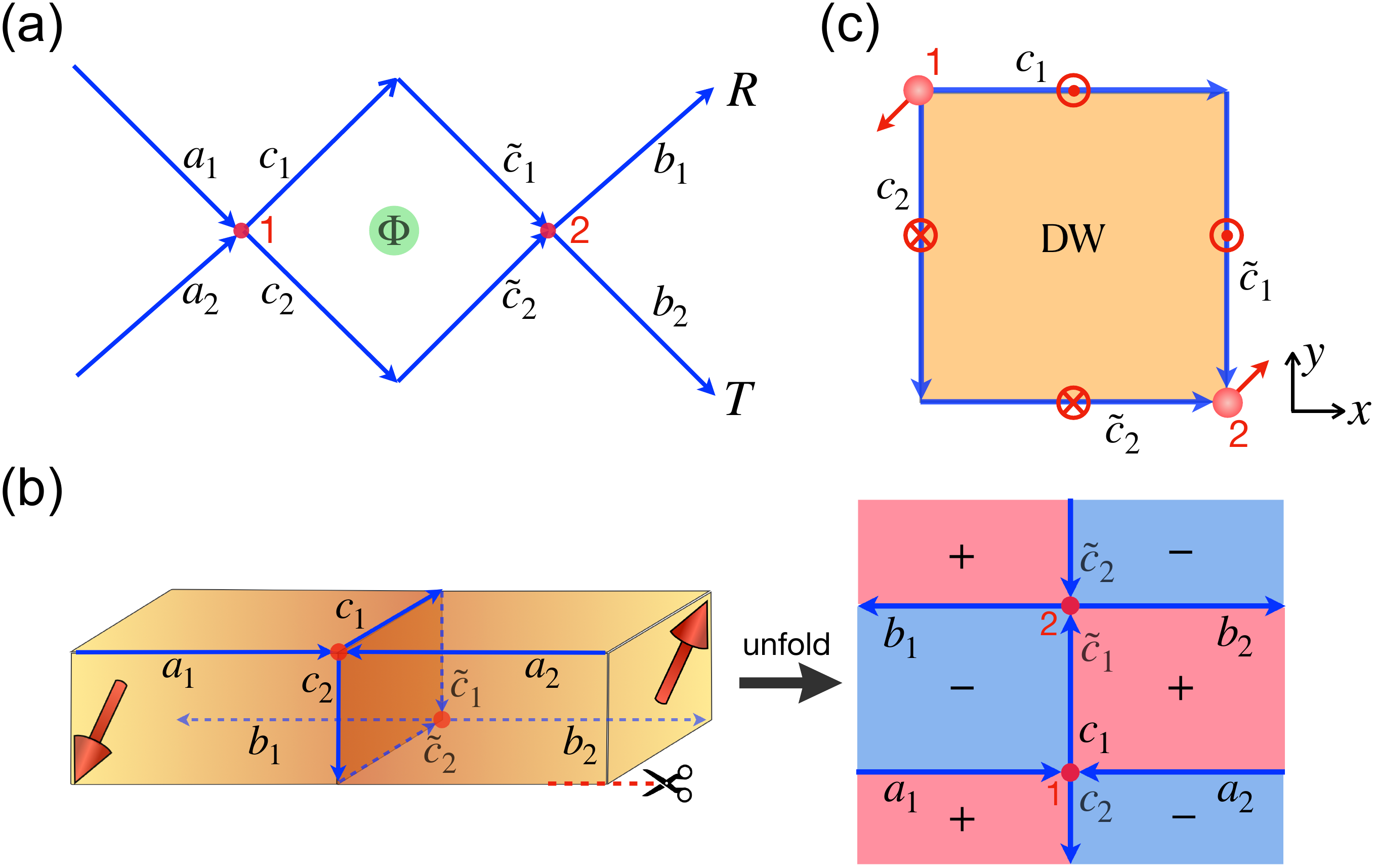} 
\caption{(a) Schematic illustration of the propagation paths of the THSs and the topological boundary states through the DW, with topologically equivalent structure to Fig. \ref{fig.SingleDWDevice}(a). The red dots labeled 1 and 2 indicate the different scattering nodes. 
(b) Unfolded view of the 3D transport device mapped onto a 2D plane (right panel), obtained by cutting along the hinge indicated in the left panel. Here the $\pm$ signs in the left panel denotes the direction of the out-of-plane magnetization.
(c) Spin texture distribution at the DW. The red arrow in the upper left (lower right) corner indicates the spin polarization of mode $a_1$ ($b_2$). The remaining four out-of-plane red arrows represent the spin polarization of the topological boundary states $c_{1(2)}$ and $\tilde{c}_{1(2)}$. Scattering nodes 1 and 2 are marked in accordance with panel (a).     }
\label{fig.ScatteringPath}
\end{figure}

\subsection{Construction of the scattering matrix}
To explain this novel AB oscillation, here we construct the scattering matrix for the THSs at the DW. To simplify the propagating paths, we extract all the conducting channels in Fig. \ref{fig.SingleDWDevice}(a), and rearrange them into a topologically equivalent structure in Fig. \ref{fig.ScatteringPath}(a). The other two topological boundary states at the DW are labeled with $\tilde{c}_1$ and $\tilde{c}_2$ [not shown in Fig. \ref{fig.SingleDWDevice}(a)]. The red dots with labels 1 and 2 denote the scattering nodes relating $a_{1(2)}$ to $c_{1(2)}$, and $\tilde{c}_{1(2)}$ to $b_{1(2)}$, respectively. Since a strict construction of the scattering matrix, demanding solving the wavefunctions of all the conducting channels and then using the wavefunction-matching conditions at the nodes, is unrealistic for such a complex geometry, we will determine the scattering matrix phenomenologically by taking advantage of the symmetry properties of the transport device. 

The scattering matrix at node 1 could be written as:
\begin{equation}
\left( \begin{matrix}
c_1 \\
c_2 
\end{matrix} \right)
=
\left( \begin{matrix}
r_1  & t_1'\\
t_1  & r_1'\\
\end{matrix} \right)
\cdot 
\left( \begin{matrix}
a_1 \\
a_2\\
\end{matrix} \right)
\equiv S_1\cdot 
\left( \begin{matrix}
a_1 \\
a_2\\
\end{matrix} \right)
\end{equation}
Similarly, the scattering matrix at node 2 could be written as:
\begin{equation}
\left( \begin{matrix}
b_1 \\
b_2 
\end{matrix} \right)
=
\left( \begin{matrix}
r_2  & t'_2\\
t_2  & r'_2\\
\end{matrix} \right)
\cdot 
\left( \begin{matrix}
\tilde{c}_1 \\
\tilde{c}_2\\
\end{matrix} \right)
\equiv
S_2
\cdot 
\left( \begin{matrix}
\tilde{c}_1 \\
\tilde{c}_2\\
\end{matrix} \right)
\end{equation}

Node 1 can be regarded as a beam splitter which scatters mode $a_1$ into modes $c_1$ and $c_2$, with probabilities $T_1$ and $T_2$, respectively, so we can write:
\begin{align}
t_1 = \sqrt{T_1} e^{i \theta_1}, \  r_1 = \sqrt{T_2} e^{i \theta_2},
\end{align}
here $\theta_{1(2)}$ are the corresponding scattering phases. 

To show the propagation paths of the THSs more clearly, we unfold the 3D transport device onto a 2D plane by cutting along the hinge shown in the left panel of Fig. \ref{fig.ScatteringPath}(b). Here the $\pm$ sign in the right panel denotes the sign of the out-of-plane magnetization. We note that scattering at node 1 obeys a two-fold rotational symmetry $C_{2{\bf n}}$ with $\bf n$ the norm of the unfolded plane. So we can set a convention that mode $a_1(c_1)$ is related to mode $a_2 (c_2)$ by this $C_{2{\bf n}}$ transformation, which yields:
\begin{align}
r_1 = r'_1, \ t_1 = t'_1.
\end{align}
The scattering matrix $S_1$ can be then written explicitly into:
\begin{align}
S_1=\left(\begin{matrix}
\sqrt{T_1}e^{i\theta_1} & \sqrt{T_2}e^{i\theta_2}\\
\sqrt{T_2}e^{i\theta_2} & \sqrt{T_1}e^{i\theta_1}
\end{matrix}\right).
\end{align}
Due to the 1D nature of all these propagating modes, the scattering at node 1 is unaffected by the finite flux $\Phi$. So we can set $\Phi = 0$ in the analysis. In this case our transport device obeys the mirror symmetry $\hat{M}_{110}$, and mode $a_1$ must be scattered into modes $c_1$ and $c_2$ with equal probabilities. This yields:
\begin{align}
T_1 = T_2.
\end{align}
The phases $\theta_1$ and $\theta_2$ are still ambiguous here. However, they are not independent. Using the unitary property of the scattering matrix:
\begin{align}
S_1 S_1^\dagger =
\left(\begin{matrix}
2 T_1 & 2 T_1 \cos{(\theta_1-\theta_2)} \\
2 T_1 \cos{(\theta_1-\theta_2)} & 2T_1
\end{matrix}\right)
= {\rm I} ,
\end{align}
where $\rm I$ is the identity matrix, we get:
\begin{align}
T_1 = T_2 = 1/2, \ {\rm and} \  \theta_1 - \theta_2 = \pm \pi/2.
\label{eq. T1/2}
\end{align}
It can be verified that the specific choice of the $\pm$ sign in Eq. (\ref{eq. T1/2}) does not affect the final value of the transmission coefficient $T$. For concreteness, we choose $\theta_1 = \pi/4$ and $\theta_2 = -\pi/4$, in which case the scattering matrix becomes:
\begin{align}
S_1=
\frac{1}{\sqrt{2}}
\left(\begin{matrix}
e^{i\pi/4} & e^{-i\pi/4}\\
e^{-i\pi/4} & e^{i\pi/4}
\end{matrix}\right)
\end{align}
Here we did not consider the spin rotation at node 1 during the unfolding process. The spin rotation will be incorporated into the transfer matrix relating $c_{1(2)}$ to $\tilde{c}_{1(2)}$, as elaborated later. The scattering at node 2 has the same configuration with node 1, as can be inferred from Fig. \ref{fig.ScatteringPath}(b), where a four-fold rotation along the norm ${\bf n}$ followed by a translational transformation relates node 1 to node 2, so we set: $S_2 =S_1$.

For transport from modes $a_1$ to $b_2$ through the DW, since the magnetization orientation flips, the spin polarization of the THSs also reverses. To account for the spin-flipping process during the propagation of THSs, we define the orbital-resolved spin polarization: 
\begin{align}
\langle s_{\alpha} \rangle_{l,n} \equiv \langle \psi_{l,n}| s_\alpha | \psi_{l,n} \rangle,
\end{align}
where $|\psi_{l,n}\rangle$ denotes the state vector at site $i$ of orbital $n$. To calculate the wavefunctions, we choose periodic boundary condition along the propagation direction of the modes, while adopting open boundary conditions for other two transverse directions. The expectation value gives the spin polarization along the $\alpha$ direction. Numerical calculations show that the spin polarizations for orbitals 1 and 2 are always opposite, i.e., $\langle s_\alpha \rangle_{l,1} = - \langle s_\alpha \rangle_{l,2}$. In Fig. \ref{fig.ScatteringPath}(c) we plot the spin polarization of orbital 2 for the THSs $a_1$, $b_2$, and the topological boundary states $c_{1(2)}$, $\tilde{c}_{1(2)}$. The results show that the spin polarization has a $\pm \pi$ rotation along the (-1, 1, 0) direction from mode $a_1$ to $b_2$, if electrons select the $c_1 \rightarrow \tilde{c}_1$ ($c_2 \rightarrow \tilde{c}_2$) path. One can also note that the distribution of the spin polarization on the DW also respects the mirror symmetry $\hat{M}_{110}$.  

In addition to the spin rotation, we also need to take into account the magnetic flux ${\Phi}$ and the dynamical phase $\theta_\lambda$. Combining all these phases together, we get the following relation for modes $c_{1(2)}$ and $\tilde{c}_{1(2)}$: 
\begin{align}
\label{eq. TM}
\left( \begin{matrix}
\tilde{c}_1 \\
\tilde{c}_2 
\end{matrix} \right)
=& 
\left( \begin{matrix}
e^{i \pi/2}  &  0\\
0  &  e^{-i\pi/2}\\
\end{matrix} \right) 
\left( \begin{matrix}
1  &  0\\
0  &  e^{-i\tilde{\Phi}}\\
\end{matrix} \right)
\left( \begin{matrix}
e^{i\theta_\lambda}  &  0\\
0  &  e^{i\theta_\lambda}\\
\end{matrix} \right)
\cdot 
\left( \begin{matrix}
c_1 \\
c_2\\
\end{matrix} \right)  \\ \nonumber
\equiv & T_M
\left( \begin{matrix}
c_1 \\
c_2\\
\end{matrix} \right),
\end{align}
where we have defined the renormalized flux $\tilde{\Phi} \equiv  \pi \Phi/\Phi_0$, and the transfer matrix $T_M$. Here the first, second, and third matrices account for the spin rotation, magnetic flux, and the dynamical phase, respectively. Since a square shape of the DW is adopted in our model, the dynamical phases $\theta_\lambda$ accumulated for path $c_1 \rightarrow \tilde{c}_1$ and path $c_2 \rightarrow \tilde{c}_2$ are chosen to be the same. 

Finally the scattering matrix, relating the incoming modes $a_1$, $a_2$, to the outgoing modes $b_1$, $b_2$, is expressed as:
\begin{align}
\label{eq. S-matrix}
S = & S_2 T_M S_1  \\ \nonumber
= &
\frac{e^{i\theta_\lambda}}{2} 
\left(
\begin{matrix}
-1 - e^{ - i \tilde{\Phi}}  & e^{i \pi/2} + e^{i( - \pi/2 - \tilde{\Phi})} \\
e^{i \pi/2} + e^{i(-\pi/2 - \tilde{\Phi})} & 1 + e^{-i \tilde{\Phi}} 
\end{matrix}
\right).
\end{align}

From this we immediately get the transmission and reflection coefficients by identifying the matrix element of $S$:
\begin{subequations}
\label{eq. TandR}
\begin{align}
T = |S_{12}|^2 = \frac{1}{2} - \frac{1}{2}\cos{(\pi \Phi/\Phi_0)}, \\
R = |S_{11}|^2 = \frac{1}{2} + \frac{1}{2}\cos{(\pi \Phi/\Phi_0)}.
\end{align}
\end{subequations}
The results are in perfect agreement with numerical calculations from the NEGF method [see Fig. \ref{fig.SingleDWDevice}(c) the red fitting curve]. The constructive (destructive) transmission at $\Phi=\Phi_0$ (0) is attributed to the $\pi$ spin rotation of the THSs traversing the DW.  One can also note that the DW structure in the 3D SOTI offers a clean and realistic platform for observing the AB phenomenon in condensed matter physics.

\section{Fabry-P{\'e}rot oscillation in the double-DW junction \label{sec: DoubleDWs}}
\begin{figure}
\includegraphics[width=8.5cm, clip=]{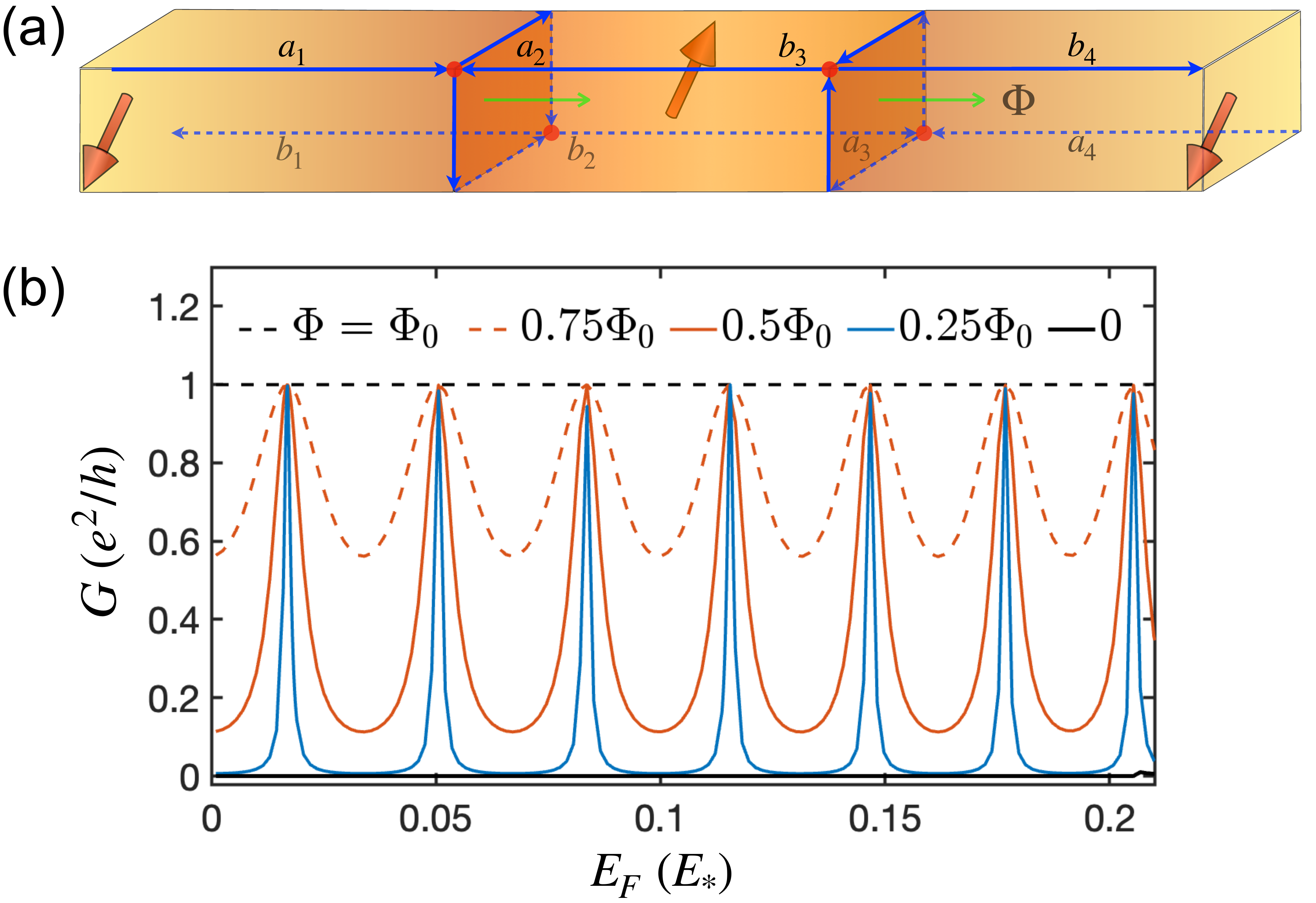} 
\caption{(a) 
Schematic diagram of the double-DW junction in a magnetic SOTI. Blue arrows indicate the propagation paths of the THSs, while 3D arrows represent the magnetization orientation along diagonal directions in the $x-y$ plane. The notations $a_{1(2)}$/$b_{1(2)}$ and $a_{3(4)}$/$b_{3(4)}$ denote the incoming/outgoing modes at the left and right DWs, respectively. Green arrows denote the applied uniform magnetic field, or equivalently, the magnetic flux threading through the DW. (b) Fermi energy dependence of the differential conductance $G$ through the double-DW junction, with different magnetic flux ${\Phi}$.  }
\label{fig.DoubleDWdevice}
\end{figure}

In this section, we consider the transport device with double DWs, where a cavity flanked by two DWs is formed, as shown in Fig. \ref{fig.DoubleDWdevice}(a). This can be realized by setting:
\begin{eqnarray}
{\bf M}({\bf r})=\left\{
\begin{array}{ll}
- M {\bf n}_{110} ,     \hspace{1mm}       & {z      <   -L_c/2  }\\
M {\bf n}_{110},       \hspace{1mm}     & {-L_c/2 \leq z     \leq L_c/2 } \\
- M {\bf n}_{110} ,     \hspace{1mm}      & {z     >  L_c/2}
\end{array} \right.
\end{eqnarray}
where the origin is set at the center of the cavity, and $L_c$ denotes the cavity length. In the discrete lattice model we use $N_c$ to characterize the cavity length, following the relation: $L_c = N_c a$. 

\begin{figure}
\includegraphics[width=8.5cm, clip=]{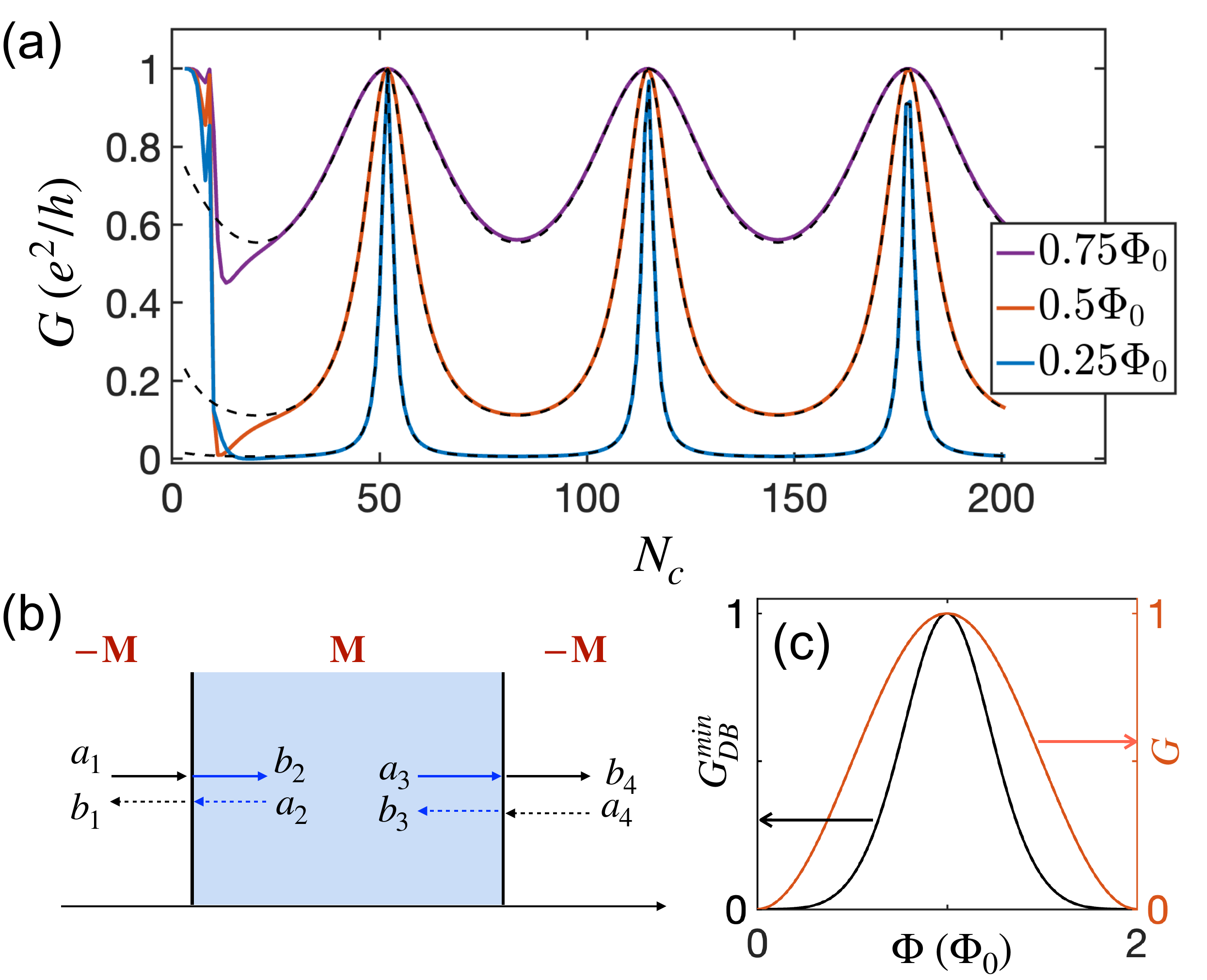} 
\caption{(a) Length $N_c$ dependence of the conductance $G$ at $E_F = 0.05 E_*$ with magnetic flux ${\Phi} =0.75\Phi_0, 0.5 \Phi_0$, and $0.25\Phi_0$. The dashed curves represent results fitted using the scattering matrix method, with the fitting parameter $\varphi_{r}$ (the phase of reflection coefficient $r$) set to 0.11  (in rad). (b) Schematic diagram of the simplified FP scattering paths to illustrate the propagation of THSs through a double-DW junction. (c) Minimum conductance $G_{DB}^{min}$ (in units of $e^2/h$) of the double-DW junction, and the conductance $G$ through a single DW, plotted as a function of magnetic flux ${\Phi}$. }
\label{fig.NcDependence}
\end{figure}

In Fig. \ref{fig.DoubleDWdevice}(b) we plot the conductance curves versus the Fermi energy $E_F$ at different magnetic flux ${\Phi}$. At ${\Phi} = \Phi_0$ we find a perfect transmission with a quantized conductance $G_{DB}=e^2/h$, and at ${\Phi} = 0$ we find a total prohibition of transmission ($G_{DB}=0$), akin to the single DW case. This is comprehensible, as electrons experience reflectionless transport across any DW at $\Phi = \Phi_0$, whereas they are completely blocked at ${\Phi} = 0$. For other values of flux, uniform FP oscillation patterns are observed. The conductance curves show resonant peaks with a quantized value $e^2/h$ at specific equidistant Fermi energies, regardless of the magnetic flux. Besides, they also exhibit minimum values $G_{DB}^{min}$ at other identical Fermi energies $E_F$, with $G_{DB}^{min}$ increasing with ${\Phi}$. Here the magnetic flux ${\Phi}$ is chosen within half of the period, e.g., $\Phi \in [0, \Phi_0]$, because the transport system at $-{\Phi}$ can be transformed from that at ${\Phi}$ by the mirror transformation $\hat{M}_{110}$, and the conductance satisfies $G_{DB}({\Phi}) = G_{DW}(- {\Phi})$. In Fig. \ref{fig.NcDependence}(a) we show the length $N_c$ dependence of $G_{DB}$ for ${\Phi} = 0.25 \Phi_0$, $0.5 \Phi_0$, and $0.75 \Phi_0$. In the short-cavity limit, $G_{DB}$ approaches $e^2/h$ due to quantum tunneling between the THSs in the two outer leads. As $N_c$ increases, $G_{DB}$ decreases sharply, and then shows similar FP oscillation patterns as Fig. \ref{fig.DoubleDWdevice}(b). 

To explain the FP oscillation, here we construct the scattering matrix for the double-DW junction. In Fig. \ref{fig.NcDependence}(b) we simplify the propagation paths of the THSs in Fig. \ref{fig.DoubleDWdevice}(a) into a 1D scattering model, where the blue region corresponds to the cavity part, and $a_{1,2,3,4} (b_{1,2,3,4})$ denote the incoming/outgoing modes. We first rewrite Eq. (\ref{eq. S-matrix}) into: 
\begin{align}
S = S_L \equiv
\left( \begin{matrix}
r & t\\
t & -r\\
\end{matrix} \right) ,
\end{align}
where we have defined the transmission and reflection coefficients $t$ and $r$ for the left DW, respectively. They should satisfy: $T = |t|^2$ and $R = |r|^2$ according to Eqs. (\ref{eq. TandR}). The scattering at the right DW can be related to the left one by an inversion transformation, which transforms modes $a_1 (b_1)$ to $a_4 (b_4)$, and modes $a_2 (b_2)$ to $a_3 (b_3)$. Thus the scattering matrix at the right DW is:
\begin{align}
S_R = 
\left( \begin{matrix}
0 & 1\\
1 & 0\\
\end{matrix} \right)
\cdot  S_L  \cdot
\left( \begin{matrix}
0 & 1\\
1 & 0\\
\end{matrix} \right)
=\left( \begin{matrix}
-r & t\\
t & r\\
\end{matrix} \right).
\end{align}
There is also a transfer matrix $T_{LR}$ relating modes $b_2, a_2$ to modes $a_3, b_3$:
\begin{align}
\left( \begin{matrix}
a_3\\
b_3
\end{matrix} \right)
= T_{LR}
\left( \begin{matrix}
b_2 \\
a_2
\end{matrix} \right),
\end{align} 
which accounts for the dynamical phase over the distance $L_c$, and can be written as:
\begin{align}
T_{LR} =
\left( \begin{matrix}
e^{ik L_c} & 0\\
0 & e^{-ik L_c}
\end{matrix}\right)
\equiv \left( \begin{matrix}
e^{i\varphi} & 0\\
0 & e^{-i\varphi}
\end{matrix}\right),
\end{align}
where $k$ is the wavenumber of THSs inside the cavity. Combining the scattering matrices $S_{L(R)}$ and the transfer matrix, we get the final transfer matrix after some straightforward algebra:
\begin{align}
\left( \begin{matrix}
b_4\\
a_4
\end{matrix} \right)
=
{ T}_{total}
\left( \begin{matrix}
a_1 \\
b_1
\end{matrix} \right),
\end{align}
with 
\begin{align}
{ T}_{total} =
\left( \begin{matrix}
\frac{(r^2 + t^2)^2}{t^2}e^{i\varphi}-\frac{r^2}{t^2}e^{-i\varphi} & \frac{-r(r^2+t^2)}{t^2}e^{i\varphi}+\frac{r}{t^2}e^{-i\varphi} \\
\frac{r(r^2+t^2)}{t^2}e^{i\varphi}-\frac{r}{t^2}e^{-i\varphi} & \frac{-r^2}{t^2}e^{i\varphi}+\frac{1}{t^2}e^{-i\varphi}
\end{matrix}\right).
\end{align}
Finally the transmission coefficient $t_{DB}$ through the double DWs can be obtained by identifying the diagonal term of the transfer matrix:
\begin{align}
\label{eq. tDW}
t_{DB}=1/{ T}_{total, 22} =
\frac{t^2 e^{i\varphi}}{1 - r^2e^{2i\varphi}}.
\end{align}
The transmission probability is thus: 
\begin{align}
T_{DB}= |t_{DB}|^2 =   \frac{T^2}{1+R^2-2R\cos{2(\varphi+\varphi_{r})}},
\label{eq. TDWfit}
\end{align}
where $\varphi_r$ is the phase of the reflection coefficient at single DW, i.e. $r = |r| e^{i \varphi_r}$. The conductance through such a double-DW junction is: $G_{DB} = \frac{e^2}{h} T_{DB}$. Due to the linear dispersion relation of the THSs, we have $\varphi = k L_c = {E}/{\hbar v} \cdot N_c a = {E}/{E_*} \cdot N_c$, so the conductance formula can be rewritten into:
\begin{align}
\label{eq. GDWfit}
G_{DB}= \frac{e^2}{h} \cdot   \frac{T^2}{1+R^2-2R\cos{2(\frac{E}{E_*} N_c +\varphi_{r})}}.
\end{align}
By setting $E=0.05 E_*$ and taking $\varphi_{r}$ as a fitting parameter, we successfully fit the conductance curves in Fig. \ref{fig.NcDependence}(a) using Eq. (\ref{eq. GDWfit}) [see the black dashed curves, where $\varphi_r=0.11$ (in rad) is used]. Again, the scatter matrix theory here perfectly explains the FP oscillation in the double-DW junction.

Eq. (\ref{eq. TDWfit}) indicates that the conductance shows a minimum value at $\varphi + \varphi_r = (2n+1)\pi$ with $n$ an integer number. This gives the minimum value: 
\begin{align}
G_{DB}^{min} = \frac{e^2}{h} \frac{T^2}{(2-T)^2} = \frac{e^2}{h} \frac{[1-\cos{(\pi{\Phi}/\Phi_0)}]^2}{[3 + \cos{(\pi{\Phi}/\Phi_0)}]^2} .
\end{align}
This is different from the single DW case where the conductance itself is tuned by the flux ${\Phi}$: $G = e^2/2h \cdot [1- \cos{(\pi{\Phi}/\Phi_0)}]$. In Fig. \ref{fig.NcDependence}(c) we plot $G_{DB}^{min}$ and $G$ as functions of ${\Phi}$ for comparison. As can be seen, both $G^{min}_{DB}$ and $G$ exhibit peaks at $\Phi=\Phi_0$, while show zero values at $\Phi=0$ or $2\pi$. Besides, the distribution of $G^{min}_{DB}$ has a Gaussian shape with its broadening narrower that $G$.

\section{Discussion and conclusion. \label{sec: Conclusion}}
The AB oscillation revealed in the single-DW junction junction is a distinctive feature of the 1D THSs. It exhibits independence on the Fermi energy, provided the energy remains within the surface gap, $\Delta_{\text{sur}} \approx 62$ meV. Additionally, we have examined the influence of DW configurations (see Appendix A) and Anderson-type disorder (see Appendix C). Our numerical results indicate that the predicted AB oscillation can be readily observed in experiments without requiring fine-tuning of system parameters.

For the double-DW junction, the same flux-controlled on-off behavior similar to the single-DW case is preserved: a perfect transmission of THSs is allowed at $\Phi = \Phi_0$ and a prohibited transmission appears at $\Phi = 0$. This conclusion can be extended to the case of multiple DWs, assuming an array of evenly spaced DWs are arranged. Hence, the magnetic flux acts as an effective control knob for switching THSs transport in magnetic SOTIs. In addition, both the AB and FP oscillations reported here serve as strong experimental signatures for identifying the presence of THSs.

In conclusion, we investigate the quantum transport of THSs across magnetic DWs in a SOTI, realized via antiparallel magnetic doping in a 3DTI. The emergence of four 1D topological boundary states at the edges of the DW forms a closed loop that connects counter-propagating THSs, allowing a magnetic-flux tunable quantum transport. Upon applying a uniform magnetic field through the DW, we observe a perfect sinusoidal AB oscillation in the two-terminal conductance. Extending the setup to a double-DW junction, where a cavity is formed, we uncover a FP oscillation. To interpret these results, we construct phenomenological scattering matrices for both the single- and double-DW cases, which accurately reproduce our numerical findings. Our study demonstrates a magnetic-flux-tunable approach for controlling quantum transport of THSs in magnetic SOTIs, offering a promising route toward the development of low-power-consumption, topological quantum electronic and spintronic devices based on HOTIs.

\section*{Acknowledgments}
This work is supported by the National Natural Science Foundation of China under Grants No. 12304070 and No. 12274466.

\section*{Data Availability} The simulation code that supports the findings of
this article are available at Ref. \cite{Code2025}.

\section*{Appendix A: Discussion on the DW configuration}
\begin{figure}
\includegraphics[width=8.5cm, clip=]{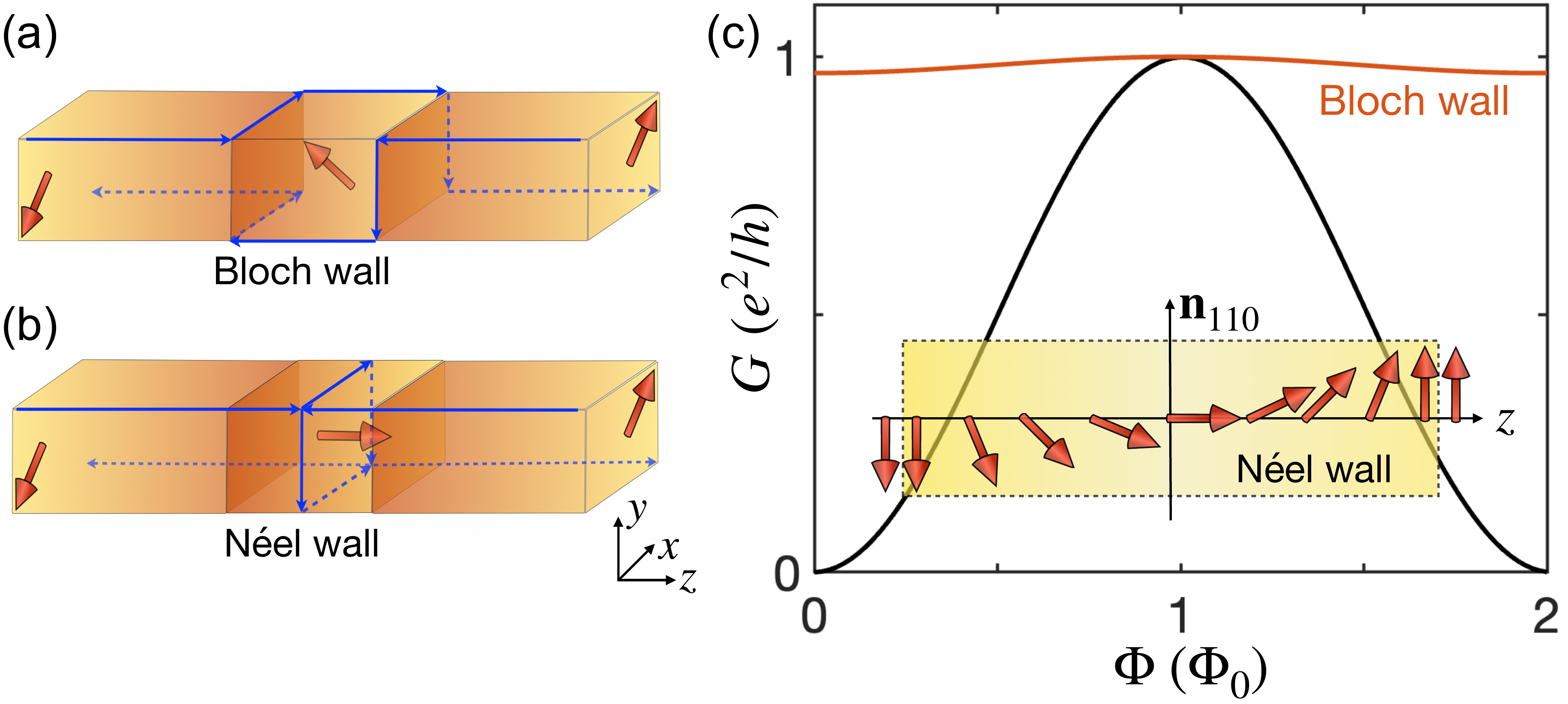} 
\caption{(a) and (b) Schematic diagrams of the Bloch wall and the N{\'e}el wall, where the magnetization rotates along the $x-y$ and the $\rm ({\bar 1} 1 0)$ planes, respectively. Here the blue arrows denote the propagation of the THSs. (c) Conductance $G$ through a Bloch(N{\'e}el) wall as a function of the magnetic flux ${\Phi}$, as shown with the red(black) curve. The inset shows the distribution of the magnetization vectors  inside the N{\'e}el wall. Here the Fermi energy $E_F$ is $0.02 E_*$ and the length of the DW (along $z$-direction) is $20a$. For the Bloch wall the magnetization follows the relation ${\bf M} ({\bf r}) = M(-(\cos{\theta}+ \sin{\theta})/\sqrt{2}, -(\cos{\theta}- \sin{\theta})/\sqrt{2}, 0)$ and for the N{\'e}el wall the magnetization follows ${\bf M} ({\bf r}) = M(-\cos{\theta}/\sqrt{2}, -\cos{\theta}/\sqrt{2}, \sin{\theta})$, where $\theta$ is a linear function of $z$, varying from zero to $\pi$ inside the DW.   }
\label{fig:NeelWall}
\end{figure}

\def\theequation{A\arabic{equation}}
\setcounter{equation}{0}
In the main text, we considered a sharp DW configuration, where the magnetization vector flips abruptly at the interface. To examine the influence of DW configuration on the AB oscillation, we now discuss two alternative and commonly encountered configurations in real materials: the Bloch wall [see Fig. \ref{fig:NeelWall}(a)] and the N{\'e}el wall [see Fig. \ref{fig:NeelWall}(b)], in which the magnetization vector rotates within the $x$–$y$ plane and the $(\bar{1} 10)$ plane, respectively. In the case of the Bloch wall, the positions of the THSs shift with changing the magnetization direction, and the counter-propagating modes can hardly hybridize. As a result, no enclosed interference loops form, and the conductance remains nearly quantized at $e^2/h$ regardless of $\Phi$, as shown by the red curve in Fig. \ref{fig:NeelWall}(c). In contrast, for the N{\'e}el wall, the counter-propagating THSs always converge at a single point where the magnetization points along the $z$-axis. This enables coherent interference, preserving the sinusoidal AB oscillation, as indicated by the black curve in Fig. \ref{fig:NeelWall}(c). Since the conductance behavior for the sharp DW and the N{\'e}el wall is essentially identical, we adopt the sharp DW configuration in the main text, and Appendix B and C below for simplicity.

\section*{Appendix B: Distribution of local density of states of topological boundary state at the DW}
\begin{figure}
\includegraphics[width=6.0cm, clip=]{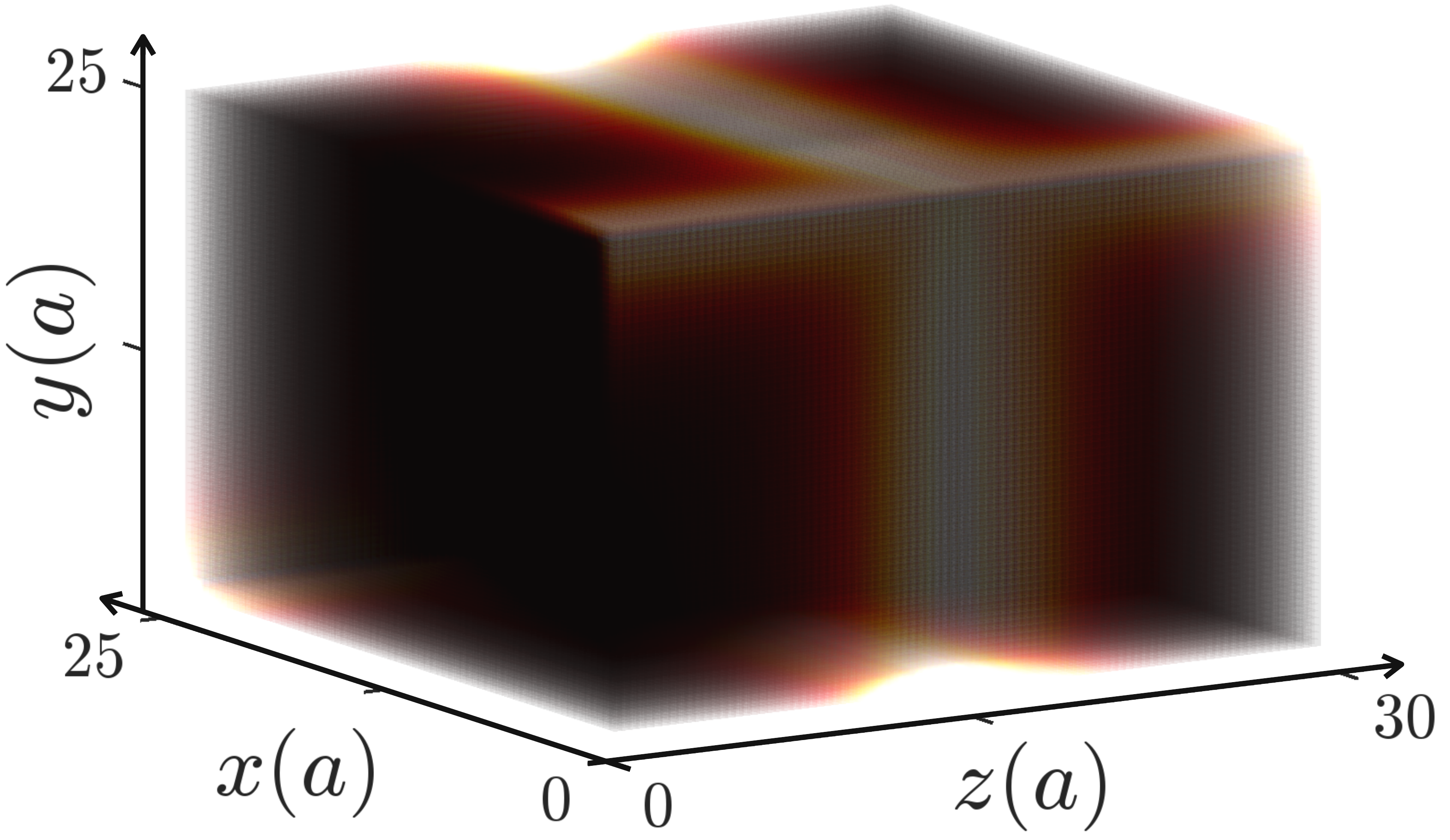} 
\caption{Distribution of the LDOSs near the DW for an infinitely-long 3DTI nanowire. Here the magnetic flux $\Phi=0$, and the energy $E=0.02E_*$. The coordinate origin has been shifted for clarity.  }
\label{fig.3DLDOS}
\end{figure}

\def\theequation{B\arabic{equation}}
\setcounter{equation}{0}
The local density of states (LDOSs) at site $m$ is calculated using the retarded Green's function introduced in Sec. \ref{sec: ModelMethods} A:
\begin{align}
\rho_{m} = - {\rm Tr} \left[ {\rm Im} G^r_c(E, {\bf r}_m) \right] / \pi,
\end{align}
where the trace is over the orbital and spin inner degrees of freedom. In Fig. \ref{fig.3DLDOS} we present the LDOSs near the DW in an infinitely-long 3DTI nanowire. The strong localization of the LDOS around the DW confirms the presence of topological boundary states, which emerge due to the reversal of the out-of-plane magnetization across the DW. In addition, the spatial distribution of the THSs are clearly visible at $(x=a, y=25a)$ and $(x=25a, y=a)$ can also be seen. These THSs are smoothly connected to the boundary states, forming continuous propagation paths mediated by the DW.

\section*{Appendix C: Effect of Anderson disorder on the conductance oscillation through the single domain wall}
\def\theequation{C\arabic{equation}}
\setcounter{equation}{0}

\begin{figure}
\includegraphics[width=6cm, clip=]{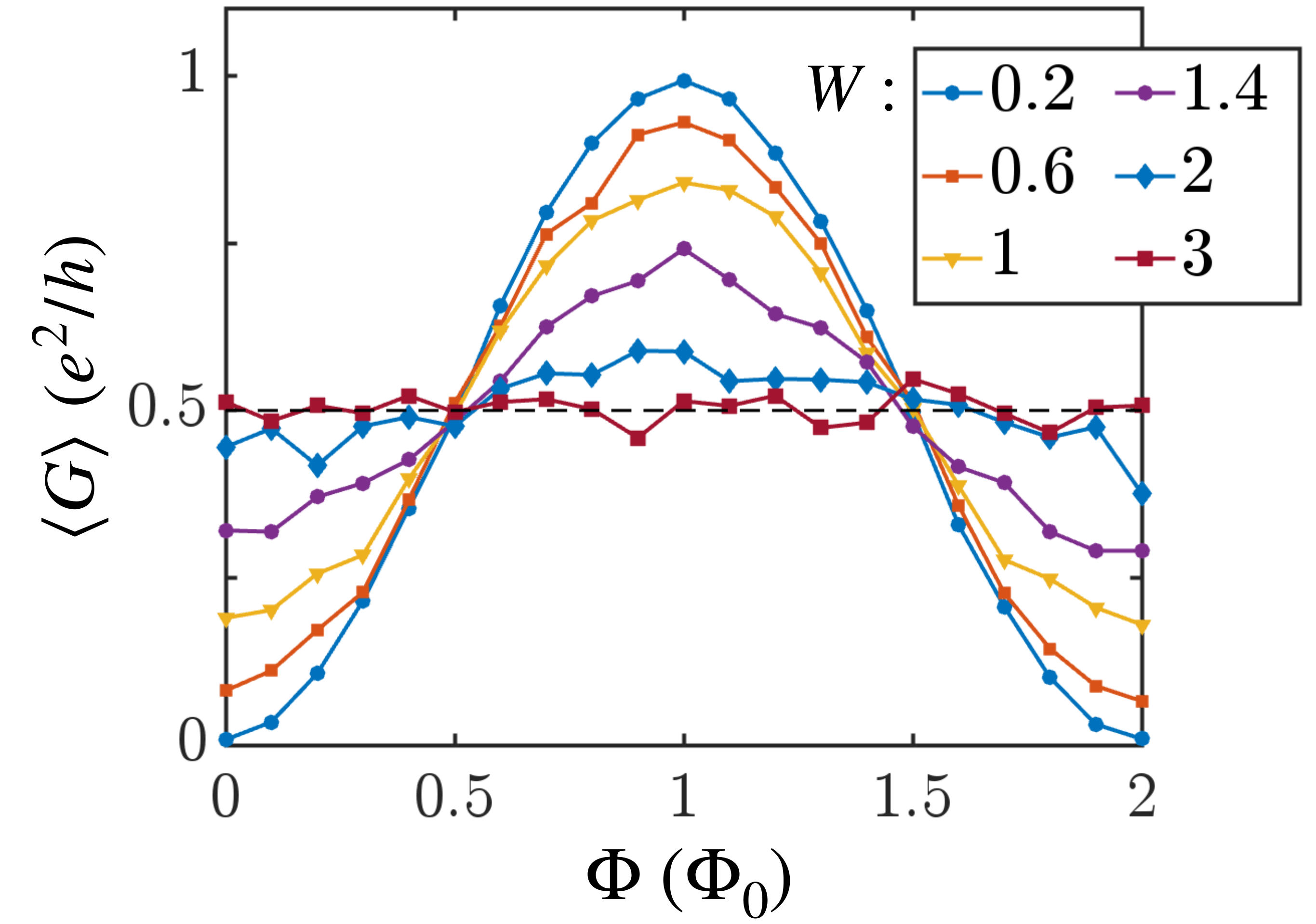} 
\caption{Ensemble-averaged conductance $\langle G\rangle$ as functions of the magnetic flux ${\Phi}$, with different disorder strength $W$ (in units of $E_*$). The length of the disordered region is set to $10 a$. The DW exists in the middle of the disordered region.   }
\label{fig:GDisorderAverage}
\end{figure}

To investigate the effect of disorder on the transport of THSs through the single DW, we consider an Anderson-type disorder existing near the DW, by adding the following term to the Hamiltonian $\hat{H}_{DW}$ in Eq. \ref{eq. HDW}:
\begin{align}
\hat{U} = \sum_{-L_{diso}/2 \leq z_l \leq L_{diso}/2} d^\dagger_l U({\bf r}_l) {\bf I}_4 d_l ,
\end{align}
where $L_{diso}$ is the length of the disordered region, ${\bf I}_4$ is the identity matrix, and $U({\bf r}_l)$ is the Anderson disorder potential at site ${\bf r}_l$, uniformly distributed in the range $[-W/2, W/2]$ with $W$ characterizing the disorder strength. Fig. \ref{fig:GDisorderAverage} shows the ensemble-averaged conductance $\langle G \rangle$ as a function of flux ${\Phi}$ under various disorder strength $W$ (in units of $E_*$), averaged over 200 disorder configurations. For weak disorder ($W=0.2 E_*$), 
the sinusoidal conductance oscillations remain largely unaffected. As $W$ increases, the oscillation amplitude diminishes, although the AB oscillation pattern is still discernible. For stronger disorder ($W \geq 2 E_*$), the oscillations are completely suppressed, and the average conductance saturates at $e^2/2h$. This saturation arises from the randomization of the dynamical phase $\theta_\lambda$ during the propagation of the topological boundary modes. An alternative explanation involves mode mixing among the four topological boundary modes \cite{LongDisorder2008, LongScaling2025}, which we leave for future investigation.

\end{document}